\documentclass[twocolumn]{aastex631}

\newcommand{\ts}{\textsuperscript}
\usepackage{hyperref}
\newcommand\jgra{\ref@jnl{J.~Geophys.~Res.~Space~Phys}}

\begin{document}

\title{On the Origin of the sudden Heliospheric Open Magnetic Flux Enhancement \\ during the 2014 Pole Reversal}

\author[0000-0002-2655-2108]{Stephan G. Heinemann}
\affiliation{Department of Physics, University of Helsinki, P.O. Box 64, 00014, Helsinki, Finland}

\correspondingauthor{S.G. Heinemann}
\email{stephan.heinemann@hmail.at}

\author[0000-0003-2061-2453]{Mathew J. Owens}
\affiliation{Space and Atmospheric Electricity Group, Department of Meteorology, University of Reading, Earley Gate, P.O. Box 243, Reading RG6 6BB, UK}

\author[0000-0003-4867-7558]{Manuela Temmer}
\affiliation{Institute of Physics, University of Graz, Universitätsplatz 5, 8010 Graz, Austria}

\author{James A. Turtle}
\affiliation{Predictive Science Inc., 9990 Mesa Rim Road, Suite 170, San Diego, CA 92121, USA}

\author[0000-0003-1662-3328]{Charles N. Arge}
\affiliation{Heliophysics Science Division, NASA Goddard Space Flight Center, Code 671, Greenbelt, MD,20771, USA}

\author[0000-0002-6038-6369]{Carl J. Henney}
\affiliation{Air Force Research Laboratory, Space Vehicles Directorate, KAFB, NM, USA}

\author[0000-0003-1175-7124]{Jens Pomoell}
\affiliation{Department of Physics, University of Helsinki, P.O. Box 64, 00014, Helsinki, Finland}

\author[0000-0002-6998-7224]{Eleanna Asvestari}
\affiliation{Department of Physics, University of Helsinki, P.O. Box 64, 00014, Helsinki, Finland}

\author[0000-0003-1662-3328]{Jon A. Linker}
\affiliation{Predictive Science Inc., 9990 Mesa Rim Road, Suite 170, San Diego, CA 92121, USA}

\author[0000-0003-1759-4354]{Cooper Downs}
\affiliation{Predictive Science Inc., 9990 Mesa Rim Road, Suite 170, San Diego, CA 92121, USA}

\author[0000-0002-2633-4290]{Ronald M. Caplan}
\affiliation{Predictive Science Inc., 9990 Mesa Rim Road, Suite 170, San Diego, CA 92121, USA}

\author[0000-0001-7662-1960]{Stefan J. Hofmeister} 
\affiliation{Leibniz-Institute for Astrophysics Potsdam, An der Sternwarte 16, 14482 Potsdam, Germany}

\author[0000-0002-5681-0526]{Camilla Scolini}
\affiliation{Institute for the Study of Earth, Oceans, and Space, University of New Hampshire, Durham, NH 03824, USA}
\affiliation{Solar–Terrestrial Centre of Excellence—SIDC, Royal Observatory of Belgium, 1180 Uccle, Belgium}

\author[0000-0001-8247-7168]{Rui F. Pinto}
\affiliation{IRAP, Université de Toulouse; UPS-OMP, CNRS; 9 Av. colonel Roche, BP 44346, F-31028 Toulouse cedex 4, France}

\author[0000-0001-9806-2485]{Maria S. Madjarska}
\affiliation{Max-Planck-Institut für Sonnensystemforschung, Justus-von-Liebig-Weg 3, 37077 G\"ottingen, Germany}
\affiliation{Space Research and Technology Institute, Bulgarian Academy of Sciences, Acad. Georgy Bonchev Str., Bl. 1, 1113, Sofia, Bulgaria}

\date{\today}

\begin{abstract}
Coronal holes are recognized as the primary sources of heliospheric open magnetic flux (OMF). However, a noticeable gap exists between in-situ measured OMF and that derived from remote sensing observations of the Sun. In this study, we investigate the OMF evolution and its connection to solar structures throughout 2014, with special emphasis on the period from September to October, where a sudden and significant OMF increase was reported. By deriving the OMF evolution at 1au, modeling it at the source surface, and analyzing solar photospheric data, we provide a comprehensive analysis of the observed phenomenon. First, we establish a strong correlation between the OMF increase and the solar magnetic field derived from a Potential Field Source Surface (PFSS) model ($cc_{\mathrm{Pearson}}=0.94$). Moreover, we find a good correlation between the OMF and the open flux derived from solar coronal holes ($cc_{\mathrm{Pearson}}=0.88$), although the coronal holes only contain $14-32\%$ of the Sun's total open flux. However, we note that while the OMF evolution correlates with coronal hole open flux, there is no correlation with the coronal hole area evolution ($cc_{\mathrm{Pearson}}=0.0$). The temporal increase in OMF correlates with the vanishing remnant magnetic field at the southern pole, caused by poleward flux circulations from the decay of numerous active regions months earlier. Additionally, our analysis suggests a potential link between the OMF enhancement and the concurrent emergence of the largest active region in solar cycle 24. In conclusion, our study provides insights into the strong increase in OMF observed during September to October 2014.
\end{abstract}
%
%


\section{Introduction} \label{sec:intro}
The Sun continuously releases magnetized plasma into heliospheric space, which is commonly known as the solar wind. The outflowing solar wind plays a key role by shaping the heliospheric medium and providing the ambient structure in which solar transients, such as coronal mass ejections (CMEs), propagate \citep[see review by][and references therein]{Temmer2021}. The solar wind can be roughly differentiated into two populations, slow and fast solar winds.\\

The sources of the slow wind are still strongly debated \citep[see][for open questions about solar wind sources]{Viall2020}{}{}. One class of theories argues that the slow wind predominantly arises quasi-statically from regions of large expansion factor at the boundaries of CHs, while other theories argue that it arises primarily from interchange reconnection \citep[see][and references therein]{Abbo2016}{}{}. This includes reconnection on top of closed loops, separatrix structures, or interchange reconnection. Regions where these processes may occur are manifold, such as photospheric magnetic field with a web-like network \citep[``S-web'';][]{2011Antiochos}, quiet-Sun regions \citep{Fisk1998}, smaller equatorial coronal holes \citep{Bale2019,Stansby2020}, coronal hole boundaries \citep[][]{wang90,Owens2018etal,Macneil2020}, edges of active regions \citep[][]{Sakao2007,Doschek2008,Harra2008}{}{}, and coronal- and pseudo-streamers \citep{Ofman2004,Riley2012}.\\

The fast wind is believed to originate in the core of the regions where the open solar magnetic field is rooted, along which the plasma is accelerated into the heliospheric space. These large-scale regions are often observed as regions of reduced emission EUV and X-ray (on-disk), white-light (off-limb) or as bright structures in He \textsc{i} 10830\AA\ \citep[see e.g.,][]{bohlin1977, 2017hofmeister,2019heinemann_catch}. The prevalence of coronal holes aligns with the solar activity cycle, where large coronal holes cover the poles in times of low solar activity and smaller equatorial coronal holes are seen during solar maximum \citep[][]{cranmer2009,Hewins2020}.  Although equatorial coronal holes can be found during solar minimum, in contrast to solar maximum they are found to feature more diffuse and fragmented boundaries \citep[``patchy'' coronal holes;][]{Heinemann2020,Samara2022}. Being regions intrinsically linked to regions of open field, coronal holes are expected to be the primary source of the heliospheric open flux throughout the solar cycle.\\

Recent studies have revealed a significant discrepancy between the heliospheric open magnetic flux derived from remote sensing (typically through potential field modeling) and the heliospheric open magnetic flux (OMF) measured in the heliosphere at 1au, which is known as the ``Open Flux Problem'' \citep{Wang1995,Wang2000,Linker2017,2019wallace}. \cite{Linker2017} found a difference of more than a factor two and \cite{Linker2021} suggested multiple sources of uncertainty in the determination of the open flux. However, even when considering the uncertainties, the missing open flux cannot be accounted for. And despite the persistent discrepancy, the trend between the in-situ measured heliospheric and the open flux based on remote sensing observations seems to be roughly correlated. \cite{Yoshida2023} suggested that the change in open solar flux is related to the evolution and interaction between the dipole and the non-dipole components of the global field. \cite{Arge2023} proposed that active regions located near the boundaries of mid-latitude coronal holes could be a possible source of the missing open flux. By considering active regions in close proximity to coronal holes as an additional source of open flux, their results show good agreement over 27 years of data with the in-situ measured OMF. However, the missing flux has yet to be definitively accounted for.\\

The evolution of the OMF seems to roughly follow the solar cycle \citep[][]{Owens2013}{}{}. Even when using improved and sophisticated algorithms to calculate the best estimate of the OMF, it yields results that raise questions about certain changes in the OMF profile, that seem not to match the global solar surface structure evolution. \cite{Frost2022} derived the evolution of the OMF including uncertainty estimates, from 1995 to 2021, and identified a sudden increase in heliospheric open magnetic flux around September -- October 2014. They find that within a few solar rotations the OMF at 1au increased by over a factor of two. This large, rapid change in OMF is a useful marker that may be exploited to connect and understand the temporal variations in photospheric, coronal and interplanetary magnetic fields. \\


In this study, we aim to trace the sudden enhancement of the OMF back to its solar origin in order to uncover its source. We derive the OMF evolution at 1au, and compare that to the modeled one at the source surface and at the solar photosphere using coronal hole observations. In addition, we examine the dynamic evolution of the solar magnetic field in 2014 in relation to emergence and decay of solar coronal holes and active regions. Thereby, we aim to shed light on solar processes that enable the opening of a large amount of magnetic flux in a short period of time relative to that of solar cycles.


\section{Methods} \label{sec:method}
To derive the evolution of the OMF and its source we investigate the magnetic field structure of the Sun by combining different techniques and data products at various heliocentric distances.

\begin{figure}
    \centering
     \includegraphics[width=0.9\linewidth]{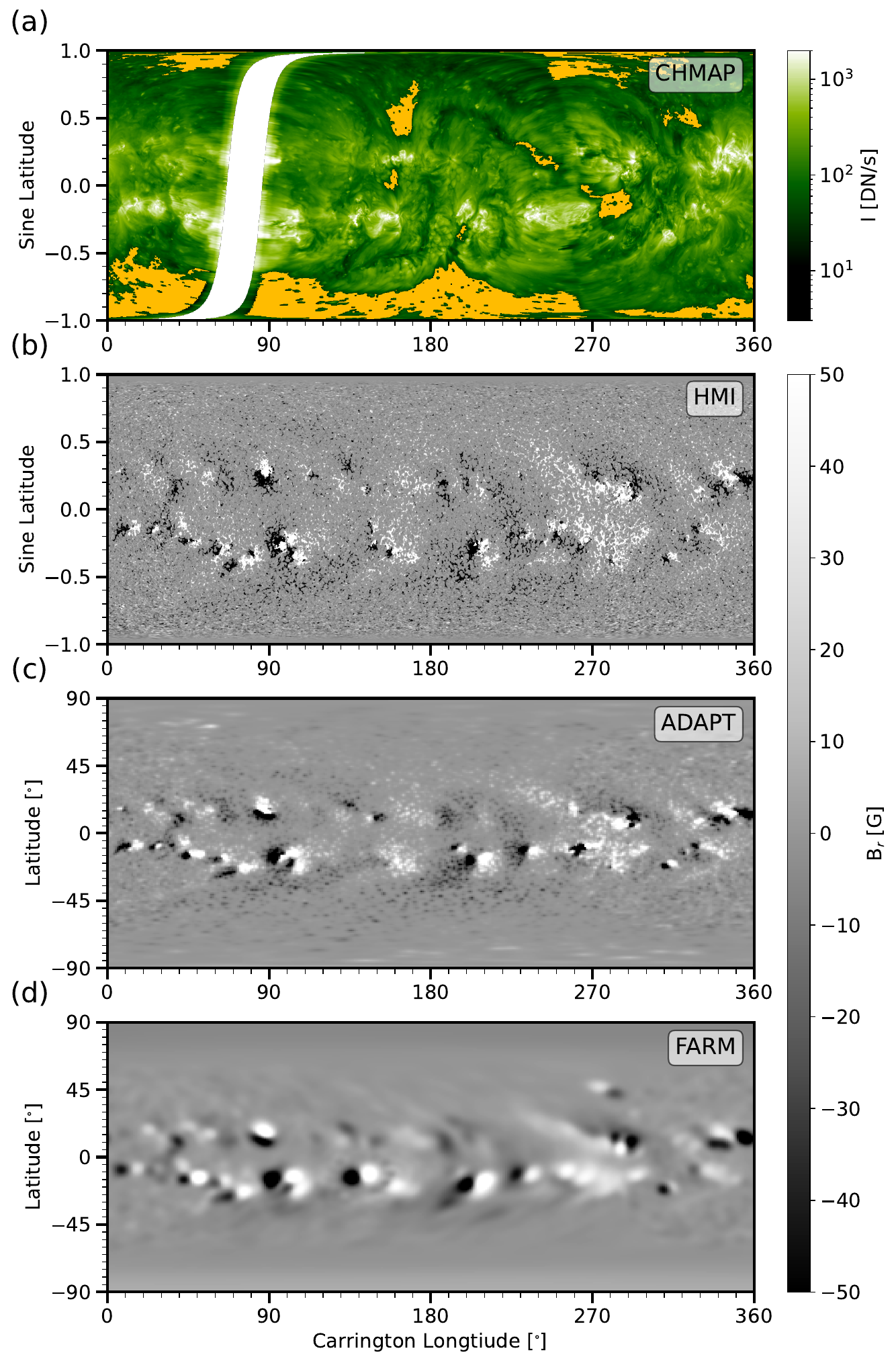}
    \caption{State of the Sun on September 19\ts{th}, 2014 as viewed from different perspectives. From top to bottom: CHMAP EUV chart with the corresponding coronal hole traction overlaid in yellow, HMI polefilled daily synoptic chart, HMI-ADAPT magnetogram and FARM magnetogram.}
    \label{fig:overview}
\end{figure}

\subsection{Coronal Holes}\label{subs:CHs}
Assuming that the majority of the open flux originates in solar coronal holes, our primary focus is to accurately determine the area and associated magnetic flux of all coronal holes from June to December 2014. To observe the full $360^{\circ}$ of the Sun, we make use of the optimal positioning of the \textit{Solar TErrestrial RElations Observatories} \citep[STEREOs;][]{2008kaiser_STEREO} and the \textit{Solar Dynamics Observatory} \citep[SDO;][]{2012pesnell_SDO}, whose combined field-of-view covers nearly the full Sun during the time period of interest. For this, we use the Coronal Hole Mapping and Analysis Pipeline (CHMAP) open-source Python software package (\url{https://github.com/predsci/CHMAP}). $195$\AA\ filtergrams from the \textit{Extreme UltraViolet Imager} \citep[EUVI;][]{Wuelser2004_EUVI}{}{} on board STEREO-A(head) and B(ehind) and $193$\AA\ filtergrams from the \textit{Atmospheric Imaging Assembly} \citep[AIA;][]{2012lemen_AIA}{}{} on board SDO are integrated in CHMAP into a single, synchronic Carrington map including automated coronal hole detections. To ensure their high quality, the individual images from each spacecraft are deconvolved, corrected for limb brightening and cross-calibrated. After the images from different instruments are cross-calibrated, the coronal hole detection algorithm is applied. An  iterative, kernel-based two-threshold scheme is used, where one (lower) intensity threshold is employed for seeding the coronal holes and a second (higher) threshold for defining the boundaries \citep[][]{Caplanetal2016}. The original detection parameters in \cite{2016caplan_LBC_IIT} were manually chosen primarily based on visual inspection and comparison to the EUV data for the period of 2012 to 2014. For this study, we used an updated and improved set of parameters to minimize the underdetection of coronal holes, at the expense of sometimes including filament channels. The extraction parameters were updated based on visual inspection and comparison to reliable coronal hole extractions of individual coronal holes \citep[using CATCH;][]{2019heinemann_catch}{}{}. As the filament channels have nearly balanced flux, these overdetections contribute very little to the open flux estimate(see Table~\ref{tab:thresh}). The ‘seeds’ are then iteratively grown until the kernel and/or upper threshold are met at all coronal hole edges. The coronal hole detection is performed on individual full-disk images and only interpolated to CR maps ($4096 \times 2048$ pixels) afterwards. This results in one map for each timestamp and each instrument. Using a minimum-intensity-merge \cite{MIDM} procedure, we combine these maps into single synchronic maps. Lastly, the map resolution was reduced by integration to $1080 \times 540$ pixels, and then further binned to a $1^{\circ}$ resolution, matching that used in the magnetic field maps. An example is shown in Figure~\ref{fig:overview}, top panel.\\ 

   \begin{table}
      \caption{CHMAP extraction parameters.}
         \label{tab:thresh}
     $$ 
         \begin{tabular}{l c c}
            \hline
            \noalign{\smallskip}
            Parameter      &  \cite{2016caplan_LBC_IIT} & This study \\
            &  $\log_{10}(I)$ & $\log_{10}(I)$ \\
            \noalign{\smallskip}
            \hline
            \noalign{\smallskip}
            seeding threshold & 0.95 & 1.05     \\
            outer threshold & 1.35 & 1.50            \\
            \noalign{\smallskip}
            \hline
         \end{tabular}
     $$ 
   \end{table}

\subsection{Surface Open Flux} \label{subs:surfaceFlux}
As the STEREO spacecrafts do not provide magnetic field remote observations, we must rely on time-dependent observations obtained solely along the Earth-Sun line to derive the magnetic field over the entire surface. To include the uncertainties in the magnetic field observations due to calibration and the ``aging-effect'' \citep[][]{HeinemannEtAl2021b}, we make use of three different magnetograms. First, we use magnetograms from the \textit{Helioseismic and Magnetic Imager} \citep[HMI;][]{2012schou_HMI} combined to the standard dataproduct \texttt{hmi.mrdailysynframe\_polfil\_720s} available via the \textit{Joint Science Operations Center} (JSOC\footnote{\href{http://jsoc.stanford.edu/}{http://jsoc.stanford.edu/}}). Secondly, for compensating, up to some extent, for the evolution of solar magnetic fields during one solar rotation, we use HMI magnetic field data assembled by the ADAPT (Air Force Data Assimilative Photospheric flux Transport) Model \citep[see \textit{e.g.,}][]{arge2013,hick2015,Barnes2023}{}{}. Note that the flux scaling method used by these ADAPT maps (B$_{\mathrm{scale}}$=$1.86576$) was factored out so that the magnetic field  values are consistent with the other magnetograms used. Lastly, we use magnetograms from the \textit{combined surface flux transport and helioseismic Far-side Active Region Model} \citep[FARM;][]{Yang2024_FARM}{}{} that features far-side active regions detected by helioseismic holography \citep[][]{Yang2023}, which are then added to an HMI-based surface flux transport model \citep[based on][]{Baumann2005}{}{} to give a more temporally accurate representation of the solar surface magnetic field. Examples of these maps are shown in the lower panels in Figure~\ref{fig:overview}. All three magnetogram data products were used at a one day cadence to match the time cadence of the coronal hole extractions. By binning to a $1^{\circ}$ resolution and co-alignment of the magnetic maps and the coronal hole observation, the coronal hole boundaries can be projected to the photosphere thus allowing to compute the signed magnetic field density ($B_{\mathrm{\textsc{ch}}}$) and to extract the signed magnetic flux ($\Phi_{\mathrm{\textsc{ch}}}$). For each day the total coronal hole flux is computed as:
\begin{equation}\label{eq:1}
    \Phi_{\odot} = \sum_i |\Phi_{\mathrm{\textsc{ch}},i}|,
\end{equation}
with $i$ subscripting every coronal hole at a given time. Note that no radial scaling to 1au is required for comparison with the total OMF derived from in-situ observations as the radial field magnitude decreases by $r^{-2}$ for an ideal Parker spiral and the area respectively increases by $r^{2}$.\\

\subsection{Source Surface Open Flux}\label{subs:SSFlux}

The open magnetic flux was calculated using a source surface potential field magnetic field model \citep[PFSS;][]{Altschuler1969} based on a finite-difference scheme. It solves a discrete Laplace equation for the scalar potential by employing a staggered representation for the magnetic field. This ensures that the magnetic field has exactly (to floating point accuracy) zero divergence and curl in the discrete representation. In addition, the input magnetogram is exactly reproduced by the model. We note that, as shown in \citet{POT3D}, open flux calculations from PFSS models are nearly resolution-independent, allowing us to use the $1^{\circ}$ map resolution reliably. Early studies \citep[\textit{e.g.,}][]{Altschuler1969,Hoeksema1984PhDT} suggested that an optimal source surface is placed around $R_{\mathrm{ss}} = 2.5 R_{\odot}$ which has then become recognized as the standard height. However, it is generally acknowledged that a fixed source surface height does not always yield reliable results but may need to be modified according to the coronal configuration. \cite{McGregor2008} varied their source surface height between $2.0$ and $2.6, R_{\text{s}}$ and \cite{Asvestari2019} used $1.4$ to $3.2, R_{\text{s}}$. This probes the coronal range in which the source surface still opens a significant amount of closed loops (lower boundary) and only opens large-scale open flux concentrations (upper boundary). Our six source surface heights were chosen to map this parameter domain. However, \cite{Asvestari2019} showed that the agreement between observed coronal hole areas and PFSS open field is strongly dependent on the choice of the source surface height for individual coronal holes. For our study, we chose source surface heights to cover this parameter range. Per definition, a lower source surface height leads to an increase in the modeled open flux. To account for this, we calculate the model flux using all three different magnetograms and employing 6 different source surface heights set at $R_{\mathrm{ss}} =[1.6,1.9,2.2,2.5,2.8,3.1] R_{\odot}$ to constrain the derived open flux at very low and very high source surface heights. Thus, for every model realization, we compute the open flux at the source surface as:
\begin{equation}\label{eq:2}
    \Phi_{\mathrm{\textsc{pfss}},R_{\mathrm{ss}}} = \sum_j |\Phi_{R_{\mathrm{ss}},j}|,
\end{equation}
where $|\Phi_{R_{\mathrm{ss}},j}|$ denotes every pixel at a given source surface. To easily demonstrate the trend over time, we derive a single time series computed from the mean flux from the 6 derived fluxes with the upper and lower boundaries constrained by the highest and lowest source surface ($R_{\mathrm{ss}} =1.6 \mathrm{~and~} 3.1 R_{\odot}$ respectively):
\begin{equation} \label{eq:3}
    \Phi_{\mathrm{\textsc{pfss}}} = \frac{1}{6} \sum^{3.1}_{R_{\mathrm{ss}}=1.6} \Phi_{R_{\mathrm{ss}}}.
\end{equation}
Even though we calculate an average of the PFSS curves for the open flux that cover the full range of commonly used source surface values, it can be used to derive trends but might not be reliable for individual events. For such events, the specific results should be examined.

\subsection{Heliospheric Open Flux} \label{subs:OMF}

\begin{figure}
    \centering
     \includegraphics[width=0.9\linewidth]{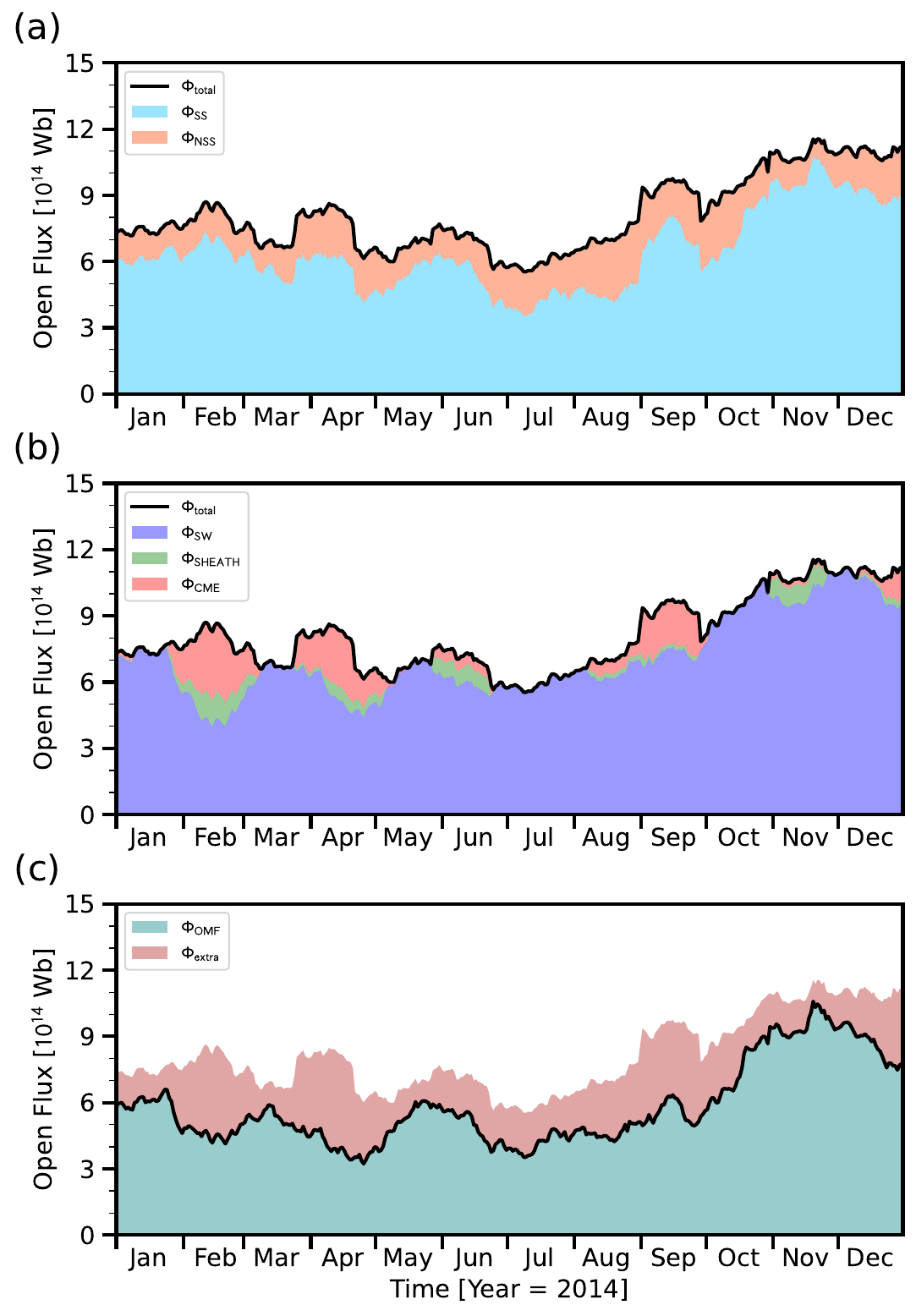}
    \caption{Estimated OMF from in-situ measurements for 2014. Panel (a) shows the contributions of source surface connected (blue) and locally inverted flux (orange). In panel (b) the contributions of ambient solar wind (SW; blue), CME sheath region (green) and CME magnetic flux rope (red) are shown. And panel (c) details the part of the OMF that is connected to the Sun and has no CME contribution ($\Phi_{\mathrm{OMF}}$; teal) as well as the contribution from ``extra'' flux (\textit{i.e.,} locally inverted and CME flux; red). }
    \label{fig:omf2014}
\end{figure}

From magnetic field data taken by the Ulysses spacecraft, \cite{Suess1996} and \cite{Suess1996a} showed that the magnitude of heliospheric magnetic field ($|\mathrm{B}_{r}|$) is independent of latitude. With this assumption, the longitudinal coverage can be obtained by integrating the in-situ measured total magnetic field over one solar rotation and further the total OMF can be calculated as $\Phi_{OMF} = 4\pi R^{2} \langle|\mathrm{B}_{r}|\rangle$, where $R$ is the respective solar distance of the observations and $\langle|\mathrm{B}_{r}|\rangle$ the 27 day average. However, this method has uncertainities related to the time resolution over which the modulus is computed. It affects how effectively small-scale inversions in B$_r$ flux (also called ``folded flux'') are averaged out \citep{Owens2017_strahl}. Choosing the averaging interval to be too short will lead to the inclusion of locally-inverted B$_r$ structures at 1au which are not present at the source surface. Too long of an averaging interval will cancel B$_r$ across the heliospheric current sheet and hence remove B$_r$ structures at 1au which are present at the source surface. In practice, this optimum time averaging will vary, particularly over the solar cycle. Instead, we use the suprathermal electron strahl to determine whether magnetic flux connects directly to the source surface, or whether it is locally inverted \citep{Owens2017_strahl}. \\

The OMF is derived from in-situ measurements obtained by the \textit{Advanced Composition Explorer} \citep[ACE;][]{1998stone_ACE} and \textit{WIND} \citep[][]{1995acuna_GSS} as stated by \cite{Frost2022}. \textit{Magnetic Field Monitor} \citep[MAG;][]{1998smith_MAG}{}{} data is utilized to determine the orientation of the heliospheric magnetic field, while electron data from the \textit{Solar Wind Electron, Proton, and Alpha Monitor} \citep[SWEPAM;][]{1998mccomas_SWEPAM}{}{} are used. For Wind, data from the \textit{Magnetic Field Investigation} \citep[MFI][]{Lepping1995_WindMAG}{}{} and the \textit{3DP Plasma Analyzer} \citep[][]{Lin1995_Wind3DP}{}{} are utilized. Detection of the strahl involves automatic identification of an increase in electron flux at 0 or 180-degree pitch angle, exceeding $50\%$ of the background flux (determined as the flux at 90 degrees pitch angle) in the 292 eV energy channel. If neither enhancement is present, it indicates the absence of the strahl (sometimes referred to as a heat flux dropout). In cases where both 0 and 180 strahl are detected, the interval is classified as counterstreaming (indicative of closed flux in the heliosphere) if the strahl fluxes are within a factor of two of each other. Otherwise, if one strahl flux exceeds twice the other, the strahl direction is determined by the highest flux. These criteria were established by comparing with previous counterstreaming and heat flux dropout occurrence rates reported in studies conducted visually \citep[][]{Gosling1992,Anderson2012,Skoug2000,Pagel2005}{}{}. Uncertainties are determined through cross-comparison between ACE and Wind data, as well as data gap analysis. They are computed as a percentage interpolation of the $90\%$ confidence intervals provided in \cite{Frost2022} to our dataset, aiming to establish a lower bound of the uncertainty. Over our time period, this uncertainty amounts to approximately $\pm 5\%$.\\

\cite{Frost2022} used this method to separate the calculated open flux into source surface connected ($\Phi_{\mathrm{SS}}$) and non-source surface connected ($\Phi_{\mathrm{NSS}}$) flux. Topologies are estimated by combining the strahl and heliospheric magnetic field directions, and a direct source-surface connection is assumed if the strahl is (locally) moving antisunward, while an inverted interplantary magnetic field is indicated when the strahl is (locally) moving sunward. The non-source surface connected flux encompasses structures like large-scale magnetic field kinks \citep{Kahler1996, Crooker2004}, and small-scale structures like ``switchbacks'' \citep[e.g.,][]{DudokdeWit2020, Squire2020} and waves/turbulent eddies. For 2014, we find that on average $22\%$ of the estimated flux is not connected to the Sun, with the contribution varying between $6\%$ and $40\%$ (see Fig.~\ref{fig:omf2014}a). As CMEs erupt in dynamic processes from the Sun, which are not captured in the magnetograms, and moreover might still be rooted at the Sun while propagating in heliospheric space, we want to exclude their contribution from the open flux calculation. We use the Richardson and Cane CME catalog \citep[][]{2010richardson_RC-list} to identify sheath regions and magnetic flux ropes. The magnetic flux rope intervals are removed and replaced with the 27-day average of the respective rotation. The same was done for the sheath region of the CME. The contributions of the different structures, \textit{i.e.} ambient solar wind, compressed solar wind (sheath), and CME flux rope, to the calculated OMF are shown in Figure~\ref{fig:omf2014}b. For the year of interest, 2014, we find that CMEs contribute up to $38\%$ of the estimated flux and $9\%$ on average. Respectively, the sheath regions contribute up to $14\%$ and on average $3.5\%$. \cite{Temmer2021a} showed that CME sheath regions are primarily made up of piled up solar wind \citep[see also][]{kilpua17b,Owens2018} and following this, we do not remove the contribution of the sheath region to the OMF. \\

By separating the different components of the derived OMF, we isolate the source surface connected contribution excluding the contribution from ICMEs, denoted as $\Phi_{\mathrm{OMF}}$ (shown in Fig.~\ref{fig:omf2014}c, teal curve). The locally inverted (i.e. non-source surface connected) and ICME related contribution varies between $8\%$ and $57\%$ and on average makes up $29\%$ of the OMF in the 2014 time period. This immediately allows us to conclude, that the strong rise around September -- October cannot be explained by such ``extra'' contributions. \\

To evaluate whether the values  derived for the year 2014 are representative in comparison to a longer time range, we extend the OMF calculations to the time period 1995--2022 covering solar cycles 23 and 24. We find that on average over these 27 years $26\%$ of the OMF is not connected to the Sun. This contribution varies between $3\%$ and $57\%$. The magnetic structure of CMEs makes up $7\%$ but can contribute up to $56\%$. During 1999--2003 solar maximum, CMEs contributed up to $56\%$ and on average $14\%$ and the following 2012--2016 maximum shows a lower contribution with up to $38\%$ and $9\%$ on average. In contrast, during times of low solar activity, the average contribution of CMEs to the OMF is significantly lower at a value of $5\%$. Summarizing, we find that the non-source surface connected and CME related contributions vary between $5\%$ and $77\%$ and on average make up $31\%$ of the OMF. This total percentual contribution of ``extra'' flux is more or less constant and not varying significantly with activity but the intrinsic variation can be relatively high.\\

\section{Results} \label{sec:res}

We analyzed the evolution of the solar magnetic field in 2014 to unravel the source of the steep increase in OMF around September -- October of the same year. By combining multiple data and processing techniques, we obtained the following results.       

\subsection{Timelines} \label{subs:timelines}

\begin{figure}
    \centering
     \includegraphics[width=0.8\linewidth]{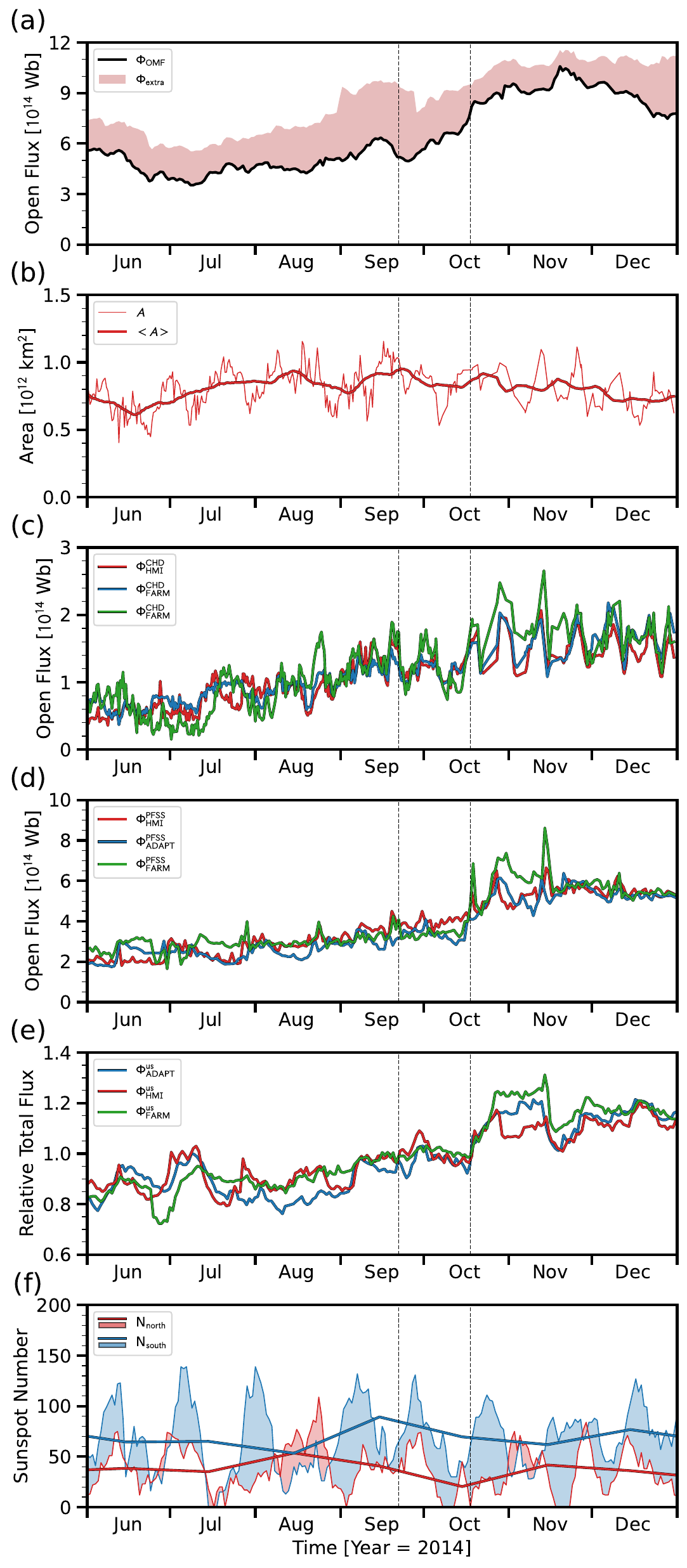}
    \caption{Evolution of the solar magnetic flux and related properties in 2014. Panel (a) shows the OMF, with the ``extra'' non-source-surface flux marked as the red shaded area. Panels (b) and (c) show the coronal hole area and the corresponding projected total open (signed) flux, respectively. In panel (d) the average PFSS calculated total open flux (from six source surface heights) is given followed by panel (e) which shows the total unsigned magnetogram flux relative to the yearly mean. Lastly, in panel (f) the daily hemispheric sunspot number is presented with the monthly sunspot number overlaid. Panels (c)-(e) show the results as function of magnetogram used (red: HMI, blue: ADAPT, green: FARM). The vertical dashed black lines highlight the first emergence and the subsequent rotation into Earth's field-of-view of the largest active region in solar cycle 24 on around September 22\ts{nd} and October 18\ts{th} respectively.}
    \label{fig:timelines}
\end{figure}

\begin{figure}
    \centering
     \includegraphics[width=0.9\linewidth]{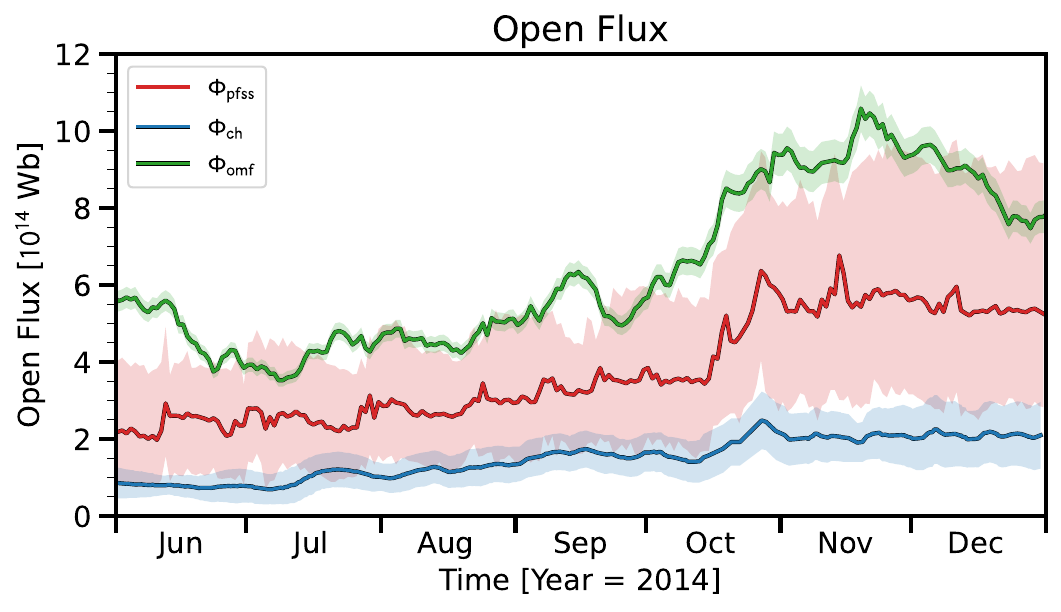}
    \caption{Open flux from different sources as function of time. The in-situ calculated heliospheric open flux is shown in green. The average PFSS open flux calculated from the three different magnetograms and 6 source surface heights is shown in red. The shaded area shows the uncertainty range, where the upper and lower boundaries are the flux calculated with R$_{\mathrm{ss}} = 3.1$ and $1.5$ respectively. The average open flux calculated from coronal hole observation overlaid on the three magnetograms is shown in blue, with the standard deviation between these three showing the uncertainty range due to different magnetograms (blue shaded area).  }
    \label{fig:timelines_res}
\end{figure}

In Figure~\ref{fig:timelines} we present the OMF, detected coronal hole areas, open flux derived from the coronal holes, open flux derived using PFSS modeling, total unsigned flux from the used magnetograms and the hemispheric sunspot numbers \citep[Source: WDC-SILSO, Royal Observatory of Belgium, Brussels;][]{sidc}. We focus on the relevant time period between June, where the open magnetic flux has its minimum in 2014 (irrespective of the method of estimation), and December, which is after the peak in the OMF. The computed magnetic flux profiles using the various magnetograms agree well and demonstrate the following behavior. In the OMF we find a slow rising trend between June and September from $4$ to $5 \cdot 10^{14}$~Wb, followed by a rise to $10 \cdot 10^{14}$~Wb within less than 2 months (Fig.~\ref{fig:timelines}a). In Figure~\ref{fig:timelines}b the CHMAP coronal hole area as a function of time is shown. Before September, we find that the area rises from around $7$ to $10 \cdot 10^{11}~\mathrm{km}^2$ from June to mid July thereafter remaining mostly constant around $8-9 \cdot 10^{11}~\mathrm{km}^2$ before eventually slowly decreasing to around $7 \cdot 10^{11}~\mathrm{km}^2$ in December. The variation in the coronal hole area is not correlated with the rise in the OMF ($cc_{\mathrm{Pearson}} = 0.00$ CI$_{95\%} = [-0.14,0.14]$) between June and October. However, it should be noted that the total coronal hole area was found to be dependent on resolution \citep[][]{POT3D}{}{}, which may alter the correlation. When considering the coronal hole open flux instead (Fig.~\ref{fig:timelines}c), we find a continuous rise in open flux between  $6 \cdot 10^{13}$~Wb in June to up to $2 \cdot 10^{14}$~Wb in November. The rise is much more continuous than the sudden jump in the OMF (although correlated $cc_{\mathrm{Pearson}} = 0.88$ CI$_{95\%} = [0.85,0.91]$) and coincides with the continuous growth of a large southern near-polar coronal hole whose signed flux increased by a factor of $3-4$ from $\approx 3 \cdot 10^{13}$~Wb to $\approx 1.1 \cdot 10^{14}$~Wb in the same time period. This coronal hole contains up to $50\%$ of the open flux originating from the observed coronal holes. \\

Figure~\ref{fig:timelines}d shows the total source surface open flux as calculated using a PFSS model. Although the average values (mean of the previously mentioned six different source surface solutions, see Eq.~\ref{eq:3}) are about a factor of two lower than the OMF \citep[see][for information on the \textit{Open Flux Problem}]{Linker2017}, the temporal evolution shows a well-correlated trend ($cc_{\mathrm{Pearson}} = 0.94$ CI$_{95\%} = [0.92,0.95]$) with the OMF. We find a slow rise between June and September from $2.5$ to $3.5 \cdot 10^{14}$~Wb, followed by a jump to $5 \cdot 10^{14}$~Wb within a month. In Figure~\ref{fig:timelines_res}, we show that even when lowering the source surface height in the model to an extremely low $1.5~$R$_{\odot}$ \citep[][]{Asvestari2019}, the OMF magnitude can barely be explained. Figure~\ref{fig:timelines}e shows the total unsigned magnetic flux of the synoptic magnetograms. Note that the total unsigned magnetic flux is highly dependent on resolution, and thus calculated for the same resolution and plotted relative to its mean value over the time period. We find that the total unsigned magnetic field increases continuously from July to October by roughly $20\%$, before rapidly increasing by another $20-25\%$ at the beginning of October. This trend is strongly correlated with the open flux calculated from the PFSS model ($cc_{\mathrm{Pearson}} = 0.94$ CI$_{95\%} = [0.93,0.96]$) as well as from coronal holes ($cc_{\mathrm{Pearson}} = 0.88$ CI$_{95\%} = [0.85,0.91]$). The correlation with the OMF is also high ($cc_{\mathrm{Pearson}} = 0.87$ CI$_{95\%} = [0.84,0.90]$).  Lastly, in Figure~\ref{fig:timelines}f the hemispheric sunspot number shows an increase in activity at the beginning of 2014, while an overall decrease is observed during the Sep-Oct period in 2014. This is seemingly in contrast to the fact that the largest active region in solar cycle 24 emerged around September 2014. Additionally, we note a slight time lag of 2-3 days between the in-situ measured and remote sensing observed open magnetic fluxes. However, this does not significantly impact the correlation coefficient in this study.\\

The onset of the steep rise in the OMF coincides with the first appearance of the largest active region observed in solar cycle 24 (NOAA identifier number 12192) on September 22\ts{nd} in the southern hemisphere, which significantly grew until it rotated into Earth's field-of-view one rotation later on  October 18\ts{th}. The second appearance coincides with the large jump in magnetic flux observed in the open flux calculated from coronal holes, from the PFSS model as well as with the unsigned total magnetic flux calculated from the magnetograms (Fig.~\ref{fig:timelines}a-e). During its disk passage on October 18 -- 30, there is only one eruptive event and CME reported from that active region which happened on October 24 \cite[see][]{Thalmann2015}.\\

In Figure~\ref{fig:timelines_res}, the open magnetic flux over time calculated from the coronal holes, the PFSS model and in-situ observations is shown. There is a large difference in magnitude between the results, however, the general trend seems to agree even though the onset and following strength of the rise varies. Regardless, all three results show a plateau-like maximum starting from mid-October. Note that the OMF is calculated within 27-day intervals, and as a result, the observed 27-day periodicity is introduced through the utilization of a rolling window. Interestingly, and in contrast to the general understanding of the origin of the open flux, we find that the contribution of open flux that has its origin in coronal holes is only $14-32 \%$ of the open flux measured in the heliosphere and $26-55 \%$ of the average open flux derived from PFSS (further depending on source surface height). The PFSS derived open flux is about $60\%$ (between $36-76\%$) of the in-situ measured one \citep[in agreement with][]{Linker2017}. 



\subsection{Complexity of Solar Evolution} \label{subs:dyn}
\begin{figure*}
    \centering
     \includegraphics[width=0.9\linewidth]{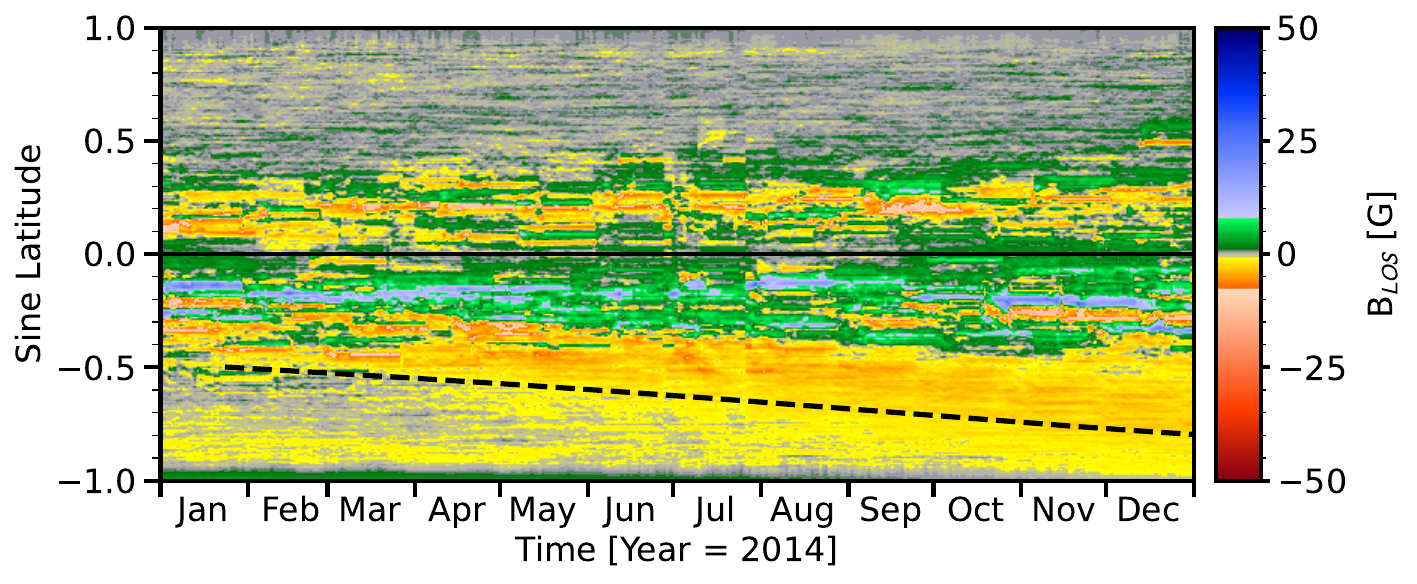}
    \caption{Longitudinally averaged magnetic field from daily HMI charts as function of time. Each column on the time axis represents the longitudinal mean as function of latitude. The black dashed line represents a plasma or flux parcel moving polewards due to solar meridional flows.}
    \label{fig:stackplot}
\end{figure*}

In the previous section, we explored how the Sun's global magnetic field properties evolved over time.  Changes in the spatial distribution of the photospheric field, however, may not be discerned by reducing the state down to a single value. To evaluate the possible role of such spatial changes, Figure~\ref{fig:stackplot} shows the longitudinal mean magnetic field from daily HMI charts as a function of time. Below $30^{\circ}$ latitude ($\sin(\lambda_{30^{\circ}})=0.5$) strong magnetic fields associated with active regions are visible. In the southern hemisphere, starting around April -- May, a poleward motion of negative polarity flux can be observed. This poleward motion of flux seems to be in the order of -- if not slightly faster than -- the meridional surface flows \citep[][meridional flow profile derived from helioseismology that takes inflows around active regions into account]{liang_18MC}. The southern polar field still features a positive remnant field from the previous cycle, crowned by an accumulation of negative flux, which observationally coincides with a large near-polar coronal hole.  Due to the opposite polarity flux migration to the pole and the associated increase in flux of the coronal hole, the remnant polar field decreases and seems to vanish around September -- October. This is in good temporal agreement with the jump in the measured OMF. Note that the observed polar fields are notoriously unreliable due to the large line-of-sight projections.
The southern polar magnetic field (defined as the mean field from latitudes $< -60^{\circ}$) is increasing in magnitude from $\approx -0.1$~G in June to $>-1$~G in December. The temporal evolution is correlated with the evolution of the OMF ($cc_{\mathrm{Pearson}} = 0.84$ CI$_{95\%} = [0.79,0.88]$) during this time period. The northern pole shows no significant change in magnetic field density.\\

\begin{figure}
    \centering
     \includegraphics[width=0.9\linewidth]{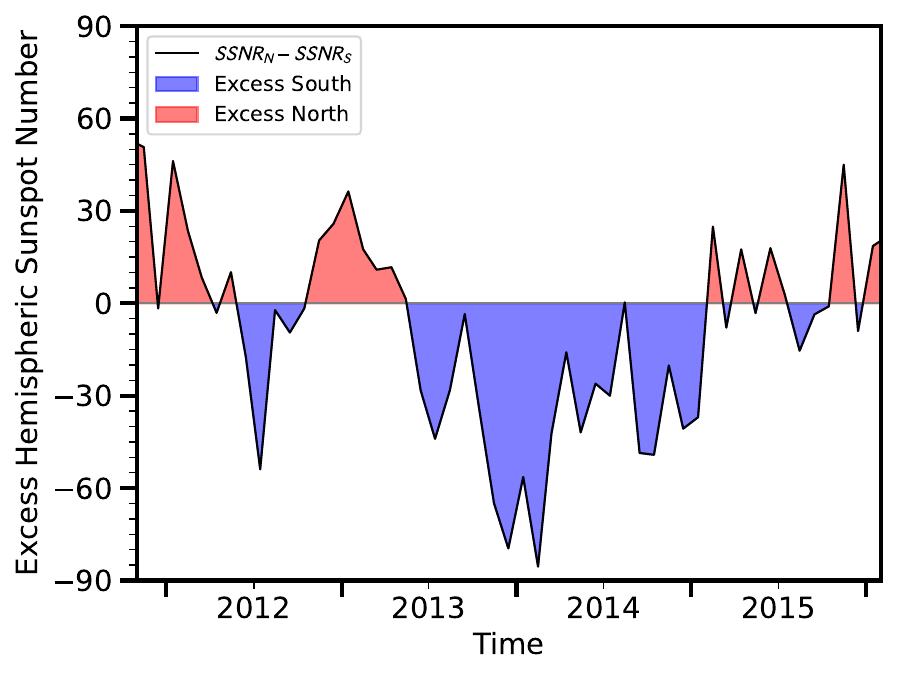}
    \caption{Excess hemispheric sunspot number over time for 2012-2016.}
    \label{fig:excessSSNR}
\end{figure}

The poleward motion of the remnant magnetic field from decaying active regions is also in agreement with the observed hemispheric sunspot number (Fig.~\ref{fig:excessSSNR}). Around January 2014, a large disparity between the southern and northern hemispheric sunspot number can be observed, with the former exceeding the latter by nearly 90, or more than a factor 3. This strong asymmetry then decreases along with the total sunspot number, which indicates that a lot of active regions in the southern hemisphere decayed during that time. This is also observed in EUV (not shown here).

\section{Discussion and Summary} \label{sec:sum}

\begin{figure}
    \centering
     \includegraphics[width=0.9\linewidth]{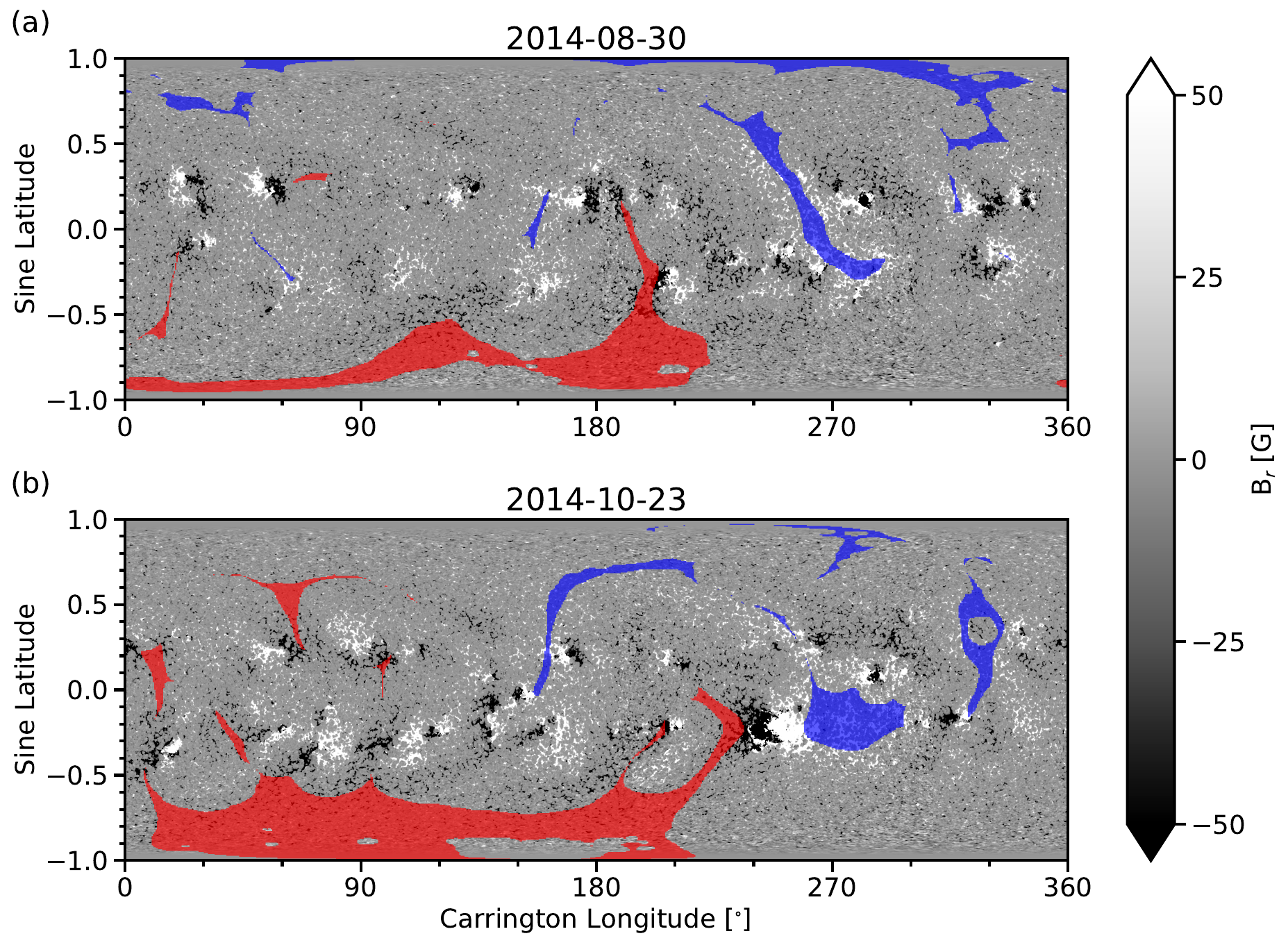}
    \caption{HMI synoptic charts from August 30\ts{th} and October 22\ts{nd} with open field footpoints overlaid. The PFSS calculations were done at a resolution of $180\times360$ pixels, at a source surface height of $R_{\mathrm{SS}}=2.1$R$_{\odot}$, and the field lines were traced to a lower boundary of R$=1.03$R$_{\odot}$.}
    \label{fig:open-close}
\end{figure}

In this study, we showed that the sudden rise in OMF is correlated with the evolution of the open flux calculated and modeled from remote sensing observations. It temporally coincides with the decay of the remnant polarity in the southern hemisphere and the consequent increase in the open magnetic flux of the southern polar coronal hole. \cite{Yoshida2023} found that the peak in the OMF is synchronous with a sudden enhancement of equatorial dipole flux derived from a PFSS extrapolation model, which is in agreement with decaying active regions moving polewards.\\

The southern active region with NOAA number 12192, was the largest active region observed in solar cycle 24 and the source of 6 X-class flares and more than 20 M-class flares in October 2014 \citep[][]{Sarkar2018}{}{}. Of the X-class flares, 4 were found to be confined \citep[][]{Thalmann2015,Baumgartner2018}{}{}. From its first appearance in September, it grew in size over six times to its appearance at the next solar rotation \citep[][]{Sarkar2018b}{}{}, which coincides with the largest jump in the observed OMF as well as in the flux derived from remote sensing observations. The emergence of this active region seemed to have caused additional flux to open. Figure~\ref{fig:open-close} shows the changes in the open magnetic field structure between October 23\ts{rd}, when the active region was at its largest and in the center of Earth's field-of-view, and two rotations prior, when the active region had not yet emerged. Although there was no significant reconfiguration of the open field structure, an additional $41\%$ of the open field area was observed, which increased the flux by over $50\%$. This is concurrent with a $25\%$ increase in the total unsigned magnetogram flux in just 2 solar rotations (for the given HMI magnetogram at a resolution of $180\times360$ pixels and a source surface height of $R_{\mathrm{SS}}=2.1$R$_{\odot}$). \\

Note that magnetograms are subject to scaling and inter-calibration to account for differences in observation technique, resolution, sensitivity, Stokes inversion and processing of the available data. For example, correction factors of B$_{\mathrm{scale}}$=$1.866$ and B$_{\mathrm{scale}}$=$1.35$  are used for HMI-ADAPT (depending on the time period and input data) to minimize offsets between different ADAPT maps. \cite{Pietarila2013} discovered a scaling factor between \textit{Vector Spectro-Magnetograph} (VSM) data from the \textit{Synoptic Optical Long-term Investigations of the Sun} (SOLIS) and HMI magnetograms might be a non-linear function of magnetic flux density, with values ranging from approximately 1 in regions of low magnetic flux density to about 1.5 in regions of high magnetic flux density. Therefore, the magnitudes of the magnetic flux calculated from remote sensing observations (e.g., coronal hole flux or PFSS flux in this study) are subject to considerable uncertainty that is challenging to assess. However, examining the ratios and trends can offer valuable insights into the evolution of the interplanetary magnetic field structure and its sources. \cite{Wangetal2022} attributes most of the missing open flux to the systematic underestimation of the observed magnetogram flux, contending that addressing this issue correctly can explain the \textit{open flux problem}.\\

Our results of the open flux derived from PFSS, coronal holes and in-situ measurements confirm the long-standing \textit{Open Flux Problem} \citep[][]{Linker2017}.  Notably, in deviation from the prevailing notion regarding the source of open flux, our findings reveal that the open flux originating from coronal holes accounts for just $14-32\%$ of the total open flux detected within the heliosphere and $26-55\%$ of the open flux derived from PFSS. This aligns with the following simple theoretical deliberations: Considering that the average magnetic field density or field strength of a coronal hole amounts to $3$~G \citep[][]{Bohlin1978,1989Obridko,2001Belenko}{}{}, then coronal holes have to cover more than $50\%$ of the solar surface to produce the observed $1 \cdot 10^{15}$~Wb of open flux. However, large coronal holes (that according to the prevailing notion should contain the majority of the flux) usually have a lower magnetic field density of around $2.5$~G \citep[][]{Heinemann2019,Hofmeister2019} which means that more than two-thirds of the Sun would need to be covered by coronal holes to explain the observed open flux. Please note that we treat polar coronal holes based on the statistics derived from non-polar coronal holes, potentially overlooking differences in their magnetic field density distribution. Furthermore, this argument overlooks the potential existence of small-scale coronal holes in proximity to active regions, where field strengths exceeding $20$~G could be present but might be obscured by the complexity of the active corona \citep[as suggested by][]{Wang_YiMing2019}{}{}. It appears that the presented results, along with the theoretical argument, suggest that the underdetection of what are traditionally referred to as coronal holes cannot resolve the discrepancy between the open flux observed from coronal holes and that derived from a PFSS model.

There is strong evidence that solar activity evolves differently in the northern and southern hemisphere \citep{Temmer2002,Hathaway2015,Lockwood2017} and that the coupling between the photosphere and the corona is also varying over the 22-year magnetic cycle \citep{wheatland2001}. This is supported by the findings that strong flare activity on average is delayed during odd-numbered cycles with respect to the relative sunspot number \citep{Temmer2003}. The Gnevyshev gap, observed in sunspot number records during the solar cycle maximum, signals the magnetic field reversal and appears to happen separately in both hemispheres \citep{Temmer2006}. \cite{janardhan18} found
that the global magnetic field reversal process was completed only in November 2014  for solar cycle 24 (which is also the peak of the open flux in our study), when the northern hemisphere had finished its final reversal. \cite{wangY2022} point out that a significant portion of the open magnetic flux might come from low-latitude regions relating to magnetic field that stays open in the wake of CMEs \citep[see also][]{Luhmann1998}. Likewise, a lag in the open flux with an even-odd cyclic behavior was reported by \cite[][]{Owens2021}{}{}. \\

Based on our results, we can attribute the large jump in heliospheric open magnetic flux around September to October 2014, where it increased by over a factor of two within less than two months from $5 \cdot 10^{14}$~Wb to $10 \cdot 10^{14}$~Wb, to an interaction of different factors, which can be summarized as follows:
\begin{itemize}
    \item The strong increase in the OMF is well-correlated with the PFSS-modeled magnetic field on the Sun ($cc_{\mathrm{Pearson}} = 0.94$ CI$_{95\%} = [0.92,0.95]$) and the open flux derived from solar coronal holes ($cc_{\mathrm{Pearson}} = 0.88$ CI$_{95\%} = [0.85,0.91]$). While the flux associated with coronal holes is well correlated, their total area is not.
    \item A large coronal hole near the southern pole was found, whose open flux increased by a factor of $3-4$ and that contains up to $50\%$ of the total open flux derived from coronal holes on the Sun. This finding suggests a potential link between the coronal hole and the observed increase in OMF, although further investigation is required to establish a conclusive relationship.
    \item Temporally, the sudden rise in OMF coincides with the disappearance of the remnant magnetic field at the southern pole, attributed to poleward flux circulations induced by the decay of a substantial number of active regions in the southern hemisphere a few months earlier. This temporal coincidence implies a connection between the decay of active regions and the subsequent increase in OMF.
    \item Lastly, we find that the strong jump in the OMF may be related to the concurrent emergence of the largest active region of solar cycle 24, which had its greatest extension in October 2014.
\end{itemize}
On the short (\textit{i.e.} sub-solar cycle) time scales considered in this study, the OMF estimated from in situ observations and photospheric extrapolations are highly correlated, though there is a systematic offset in the magnitudes \citep[also see][]{Linker2017}. It has previously been shown that on longer time scales (\textit{i.e.} cycle-to-cycle variations), this correlation breaks down \citep[][]{2019wallace,Frost2022}, possibly due to the systematic offset having a time-varying component. This suggests the presence of two distinct, decoupled sources of OMF, each operating on different time scales. Furthermore, a significant portion of this disparity may be attributed to the source regions of the ``missing'' open flux, which is improbable to originate from the centers of observed coronal holes. Thus, to gain a comprehensive understanding of the Sun's magnetic structure and the heliosphere, additional efforts aimed at solving the \textit{Open Flux Problem} are imperative.


%

\begin{acknowledgments}
We thank the International Space Science Institute (ISSI, Bern) for the generous support of the ISSI team “Magnetic open flux and solar wind structuring in heliospheric space” (2019--2021). The SDO and STEREO image data are available by courtesy of NASA and the respective science teams. This research was funded in whole, or in part, by the Austrian Science Fund (FWF) Erwin-Schr\"odinger fellowship J-4560. SGH expresses sincere thanks to Yi-Ming Wang for inspiring and fruitful discussions. The ADAPT model development is supported by Air Force Research Laboratory (AFRL), along with AFOSR (Air Force Office of Scientific Research) tasks 18RVCOR126 and 22RVCOR012. The views expressed are those of the authors and do not reflect the official guidance or position of the United States Government, the Department of Defense (DoD) or of the United States Air Force. MO is part funded by Science and Technology Facilities Council (STFC) grant number ST/V000497/1. MM acknowledges DFG grants WI 3211/8-1 and 3211/8-2, project number 452856778, and was partially supported by the Bulgarian National Science Fund, grant No KP-06-N44/2. JP acknowledges funding from the Academy of Finland project SWATCH (343581). RP acknowledges financial support by EU H2020 grant No 870437 and by ERC (SLOW-SOURCE–DLV-819189). EA acknowledges support from the Academy of Finland/Research Council of Finland (Academy Research Fellow grant number 355659). CNA is supported by the NASA competed Heliophysics Internal Scientist Funding Model (ISFM). RC, CD, and JAL of Predictive Science were supported by NASA grants 80NSSC22K0893, 80NSSC19K0273, and NSF grant ICER 1854790.
\end{acknowledgments}

\bibliographystyle{aasjournal}



\end{document}